\documentstyle[times,pramana,epsf,floats]{ias}
\begin{document}
\mark{{Event by event...}{WA98 Collaboration}}
\title{Event-by-event search for charge-neutral fluctuations in Pb-Pb collisions at 158 AGeV}

\author{Madan M. Aggarwal}
\address{ Department of Physics, Panjab University , Chandigarh-160014, India\\
                       ( For the WA98 Collaboration )}
\pacs{2.0}
\abstract {
Results from the analysis of data obtained from the WA98 experiment 
at the CERN
SPS have been presented. Some events have been filtered which
show photon excess in limited $\eta-\phi$ zones within the overlap 
region of the charged particle and photon multiplicity detectors.
}

\maketitle

Charged-neutral fluctuations have been one of the predicted 
signals for chiral symmetry restoration
in heavy-ion
collisions at ultra-relativistic energies Refs.[1-7].
It is hoped that the extreme energy densities
residing inside the spatial region between receding baryonic 
slabs shortly
after a heavy ion collision may provide the physical conditions necessary 
for the formation of a
chiral condensate aligned in a direction different from the true vacuum.
It has been predicted  that the chiral field relaxes to the true vacuum in 
such domains emitting coherent pions. The coherent pion emission by 
domains of such
Disoriented Chiral Condensates (DCCs)  within the collision volume
results in large fluctuations in the ratio of neutral to charged 
pions in
those domains. The observed Centauro and Anti-Centauro events in cosmic 
ray collisions [8] may be due to the formation of large
domains of DCCs. 

Recently  DCC formation has been investigated by the
Minimax experiment in  $p-\bar{p}$ collisions
at the Tevatron energy [9] and by the  
NA49 experiment [10] at the CERN SPS but no conclusive evidence 
for DCC formation was observed. The WA98 experiment has also 
studied the correlated neutral to charged particle fluctuations globally
[11] and by the Discrete Wavelet Transform (DWT) 
technique [12] in Pb-Pb collisions at 158 AGeV and 
upper limits on DCC formation have been reported.
In this paper we present first
results on an event-by-event search for charged-neutral
fluctuations in limited $\eta-\phi$ phase space regions.

Results are based on the  Pb-Pb collision data at 158 AGeV taken by 
the WA98 experiment at the CERN SPS, under zero magnetic field conditions,
during the 1996 run time. 
In the present
investigation we have used the  distribution of photons
measured by  
the Photon Multiplicity Detector (PMD) and that of charged particles 
measured by the Silicon
Pad Multiplicity Detector (SPMD).
The PMD, placed
at 21.5 meters downstream of the target, consisted
of plastic scintillator 
pads of varying sizes, arranged inside 28 box modules, placed behind 
3$X_{0}$ thick lead converter plates [13]. The energy
 deposited in each pad 
due to
preshowering was measured by a CCD camera system and was digitized. 
The clusters of neighbouring affected pads were used to count the 
number of hits on the PMD. Those clusters with ADC values greater than or equal to
three times that of the minimum ionising track were regarded as photon 
- like hits and will be referred to as photon hits (N$_{\gamma}$)
throughout this paper.
The photon counting efficiency was found to be
68\% to 73\% for central and peripheral events,respectively. 
In the present analysis,the data from 22 
central cameras covering the pseudorapidity region of 2.9 $ < \eta < 4.2$
has been used. The 
SPMD, consisted of 22 radial and 46 azimuthal segments in
each of the quadrants and covered the pseudorapidity region of 2.35 $ < \eta <
3.75$. The detection efficiency of the SPMD was 99\%.
The Midrapidity Calorimeter (MIRAC)
placed at $24$ m downstream of the target measured the flow of transverse 
energy  ($E_{T}$) emanating from the collision.
$E_{T}$ is used to characterize the event centrality.

We study the fluctuations
in the neutral pion fraction 
($f$) \[{f = \frac{N_{\pi^o}}{N_{\pi^o}+N_{\pi^{\pm}}}} \approx \frac{N_{\gamma}/2}{N_{\gamma}/2+N_{ch}}\]
 which should follow a probability distribution of the type :

\[P(f) = \frac{1}{2\sqrt{f}}\]
\\
for emission from DCC domains. 

We have confined ourselves to the study of events having {\em photon excess}
in azimuthal
patches within the overlap zone of the PMD and the SPMD. For these patches,
the purity of the photon sample observed by the PMD is higher
than those for normal patches due to the depleted flux of charged particles.

We have analysed a set of 196K events, 
corresponding to the top 15\% of the minimum bias
cross-section, as determined from the transverse energy ( $E_{T}$ ) 
measured by the MIRAC. The events are analyzed  for fluctuations in the neutral
pion fraction ( $f$ ) in the $\eta-\phi$ phase space region
on an event-by-event basis.
In this paper an attempt has been made to identify anticentauro type events.
 Such events are characterised by an excess of
gammas over the number of charge particles in certain $\eta-\phi$ domains.
Since the location of these domains is not known it is essential
to extend our search over the whole range of $\phi$ values. In our method we
choose a particular azimuthal window, $\Delta\phi$, in the pseudorapidity
region,  2.9 $ < \eta \le $ 3.75. 
We  scan the entire azimuthal range from $0^{\circ}$ to 
360$^\circ$ in order to search for a patch having maximum value of $f$ which 
will be referred to as $f_{max}$.
This scan is performed by successive $2^{\circ}$ rotations of the
($\Delta\eta$-$\Delta\phi$) patch.
To minimize the statistical fluctuations, a patch with maximum $f$ value
in an event was required to have at least 40 $\gamma$'s 
corresponding to a  15\% statistical error.

Results have been compared
with  simulated events obtained using the VENUS 4.12 event generator [14].
These
events were processed through the WA98 detector setup using GEANT 3.21.
The output of these processed simulated events is analysed in a similar
manner as data and is referred to as (V+G). The
statistical significance of the above results  is obtained by comparing the 
results with those from mixed events. Mixed events
are obtained by randomly mixing photons and charged particles independently 
while keeping the $N_{\gamma}$ - N$_{ch}$ correlation intact and also taking
proper care of the two track resolution. 

\begin{figure}
\setlength{\epsfysize}{3.5in}
\centerline{\epsffile{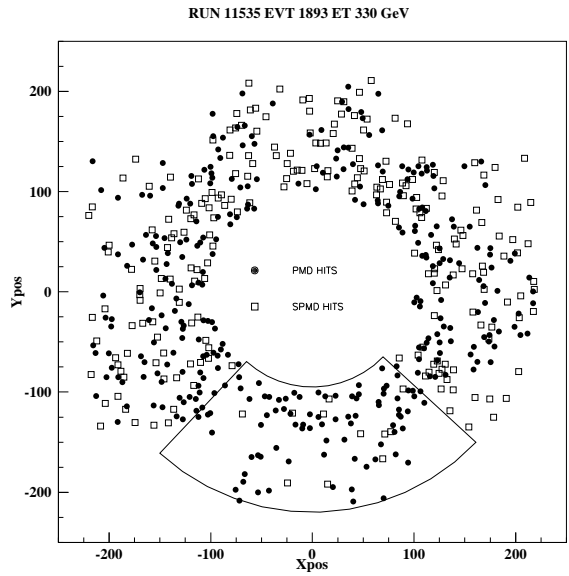}}
\caption{Plot showing photon hits (PMD) and charged particle hits (SPMD)
in an azimuthal plane. The marked 90$^{\circ}$ patch corresponds to $f_{max}$ = 0.77.}
\label{fig1}
\end{figure}

\begin{figure}
\setlength{\epsfysize}{3.5in}
\centerline{\epsffile{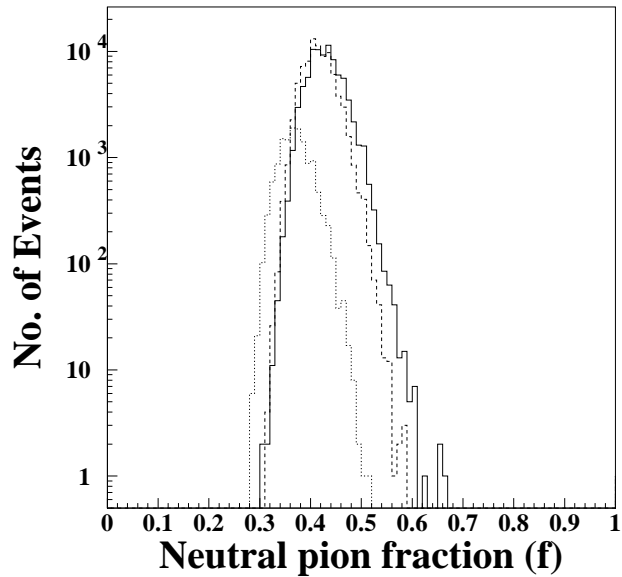}}
\caption{$f_{max}$ distributions for 60$^{\circ}$ patch 
 for Data (solid histogram) , Mixed Events (dashed histogram) and V+G 
(dotted histogram).}
\label{fig2}
\end{figure}

An event displaying PMD hits (filled circles) 
and SPMD hits (open squares) within the overlap $\eta-\phi$ zone
(obtained by using the method described above) is shown
in Figure 1. A patch of $\Delta \phi$ = 90$^{o}$ in azimuth 
is also marked.
It is seen that in this domain the number of charged particles is
only 12 as compared to 84 photons which corresponds to  
 $f_{max}$ = 0.77 in this event. Figure 2 shows  histograms of the maximum 
value of $f$, obtained in an event, for Data, Mixed Events, and V+G for a 
domain size of $\Delta \phi$ =60$^{\circ}$.
It is seen that the $f_{max}$  distribution for the data 
extends to much larger values than mixed events and (V+G) events. 
A similar trend was observed for lower centrality
 intervals (i.e., top 5 - 10 \% and top 10 - 15 \% events). This analysis
 was also carried out for
40$^{\circ}$ and 90$^{\circ}$ patch sizes. Table 1 lists the
number of events per 10 K, corresponding to $f_{max}>$0.55 for
various values of  azimuthal windows (i.e., 40$^{\circ}$, 60$^{\circ}$,and 
90$^{\circ}$) for three centrality bins with at least 40 $\gamma$'s 
in a domain. 
The results for simulated V+G events and generated mixed events for different 
centralities and patch sizes are also listed in the table for comparison.
We see that events with $f_{max}$ $>$ 0.55 are more frequent 
in data as compared to
those seen in mixed events and in V+G events. The fraction of these events
decreases significantly as we increase the patch size. Furthermore, the fraction
of these events increases significantly with decrease in centrality. 
Figure 3 exhibits the azimuthal
distribution of patches having $f_{max}$ $>$ 0.55 in data in the
 top 15 \% central 
events for a  60$^{\circ}$ patch . It is seen that these patches are more or less uniformly 
distributed in azimuth.

\begin{table}
\label{number}
\renewcommand{\thefootnote}{\fnsymbol{footnote}}
\caption[]{No. of events, per 10 K, with neutral pion fraction
$f_{max} >$0.55 in
data, VENUS+GEANT and Mixed Events for $40^\circ$, $60^\circ$, and
 $90^\circ$ patch sizes for different centrality bins selected by the MIRAC.}
\vspace{0.5cm}
\begin{minipage}[b]{5.0in}
\begin{center}
\begin{tabular}{|c|c|c|c|c|c|} \hline
Event Type & Sample & No. of events &  $\Delta\phi$=$40^\circ$&  $\Delta\phi$=$60^\circ$&  $\Delta\phi$=$90^\circ$
 \\ \hline
 & Data & 84K &185.9$\pm$4.7 & 16.8$\pm$1.4 & 0.9$\pm$0.3 \\
Top 5\% & V+G & 13K&0.7$\pm$0.7 & $\approx$0   \footnotemark \footnotetext{* The number of events have been estimated, for a similar statistics as in the data, assuming a gaussian
 distribution for $f$ as in Fig. 2 for V+G events. } & $\approx$  0$^{*}$ \\
 & EvtMix &83K& 37.8$\pm$2.1 & 1.9$\pm$0.5 & 0 \\ \hline
 & Data &78K& 325.0$\pm$6.5 & 69.3$\pm$3.0 &  2.7$\pm$0.6\\
Top 5-10\% & V+G &8K& 1.2$\pm$1.2 &   1.2$\pm$1.2 & $\approx$0 $^{*}$ \\
 & EvtMix &77K& 109.3$\pm$3.8 & 14.8$\pm$1.4 & 0.4$\pm$0.2 \\ \hline
 & Data &34K& 317.1$\pm$9.5 & 158.6$\pm$6.7 & 12.3$\pm$1.9 \\
Top 10-15\% & V+G &6K& 1.5$\pm$1.5 & 3.0$\pm$2.1 & $\approx$ 0 $^{*}$ \\
 & EvtMix &30K& 120.0$\pm$6.3 & 56.9$\pm$4.3 & 1.6$\pm$0.7 \\ \hline
\end{tabular}
\end{center}
\end{minipage}
\end{table}

\begin{figure}
\setlength{\epsfysize}{3.45in}
\centerline{\epsffile{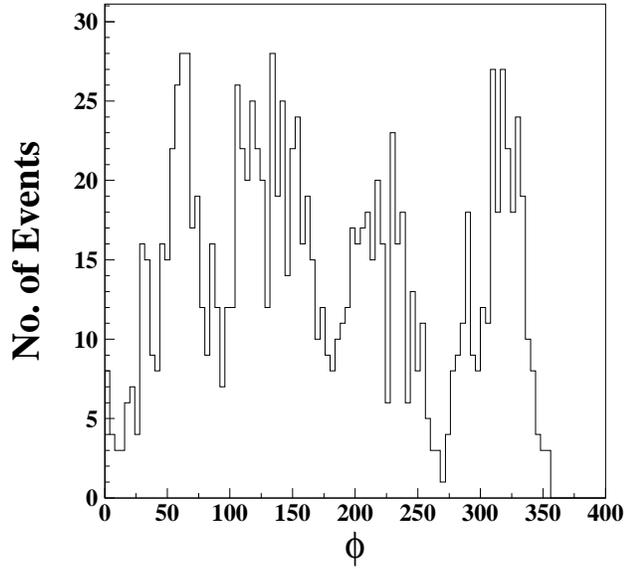}}
\caption{$\phi$ distribution of the patches having $f_{max}$ $>$0.55 for
60$^{\circ}$ domain in the top 15\% of events.}
\label{fig3}
\end{figure}
\begin{figure}
\setlength{\epsfysize}{3.45in}
\centerline{\epsffile{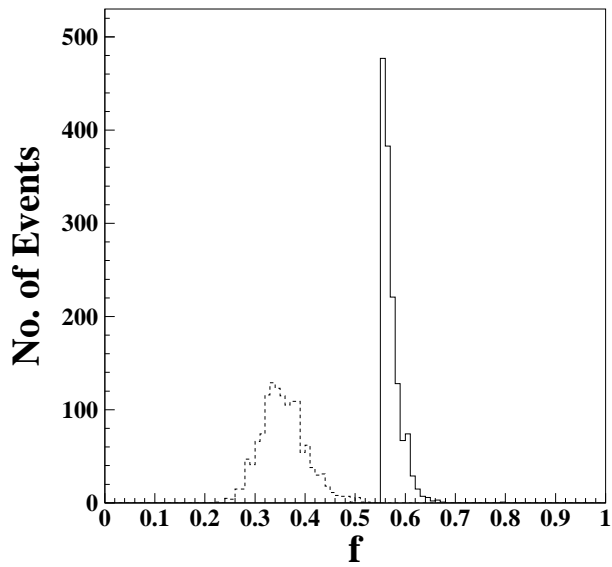}}
\caption{
 Neutral pion fraction ($f$) distributions
 for events having $f_{max}$ $>$0.55
 (solid line) and for preceding and succeeding
 events (dashed line) for
60$^{\circ}$ patch having N$_{\gamma}$  $>40$.}
\label{fig4}
\end{figure}

To test the authenticity of these events the immediate preceding
and succeeding events in the data have been checked for  the same $\eta-\phi$
patch and the neutral pion fraction ($f$) was calculated for the same
patch size. The distribution of these $f$ values is shown by the 
dashed line in Figure 4. The cut of N$_{\gamma}$ $>$ 40 has also been 
used here. In Figure 4 we also show, for comparison, the neutral 
pion fraction for patches having  $f_{max}$ $>$ 0.55  (solid line)
with N$_{\gamma}$ $>$ 40.
 It is seen that the dashed curve
has its peak around 0.35 and it shows a behaviour similar to that of generic 
events. It means that events with $f_{max}$ $>$ 0.55 are special events, having large
non-statistical charged-neutral fluctuations. In this sense, these events
closely resemble the {\em anti-centauro} type events found in  cosmic 
experiments [8].

In summary, we have found events with large  charged-neutral fluctuations
in our data which show $f_{max}$ $>$ 0.55 in certain $\Delta\phi$ domains. 
These domains are uniformly distributed in azimuth.
The fraction of such events are much larger in the data as compared
to those seen in the mixed events and V+G events. The percentage of
these events increases significantly with decreasing patch size. Also 
the fraction of such events increases significantly as we decrease
the centrality. This is the first time that {\em anticentauro}  type events
have been observed in heavy ion collisions in an accelerator based 
experiment.

\end{document}